# Mathematical and computational modeling for describing the basic behavior of free radicals and antioxidants within epithelial cells

Alvaro Juan Ojeda Garcia

**ABSTRACT**[1]

The traditional methods of the biology, based on illustrative descriptions and linear logic explanations, are discussed. This work aims to improve this approach by introducing alternative tools to describe and represent complex biological systems. Two models were developed, one mathematical and another computational, both were made in order to study the biological process between free radicals and antioxidants. Each model was used to study the same process but in different scenarios. The mathematical model was used to study the biological process in an epithelial cells culture; this model was validated with the experimental data of Anne Hanneken's research group from the Department of Molecular and Experimental Medicine, published by the journal *Investigative Ophthalmology and Visual Science* in July 2006. The computational model was used to study the same process in an individual. The model was made using C++ programming language, supported by the network theory of aging.

## I. INTRODUCTION

Traditional methods of the biology for describing and representing complex living systems have been focus of discussions in the past years (Wang, 2010). These methods are mostly illustrative and descriptive lack of mathematical explanations and become insufficient for describing highly complex processes and their relations (Cui et al, 2009). They typically contain an amazing combination of diverse elements, which present specific dynamics (Taylor et al, 2011)

This work aims to improve such methods using mathematical and computational models in order to better describe cell interactions. This study is divided in the following scenarios: a cell culture for the mathematical model and a human body for the computational C++ based simulation program. The mathematical model has been developed by applying the Lanchester's equation (Jiuyong et al, 2009). This expression simplifies the interaction between free radicals and cells protected by antioxidants as a biological combat of two forces coming across. In addition, the simulation program describes an organism involved in regular activities, which are described by a tentative schedule.

The mathematical model was contrasted with Anne Hanneken's experimental work, published by the journal "Investigative Ophthalmology and Visual Science" in July, 2006 (Anne Hanneken et al, 2006). This comparison takes into account the experimental results of exposing cells to hydrogen peroxide ($H_2O_2$) or t-butyl hydroperoxide (t-BOOH) (free radicals). The mathematical model closely follows the surviving cells function. Finally, the computational model was programmed in C++ language. This simulation calculates the dead skin cells from the body, estimated in around 30000 to 40000 every minute (Ling, 2007)

This paper is organized in five sections. Section 2 presents the theoretical background of free radicals and antioxidants. In section 3 the biological models are described. This section contains the mathematic model of free radicals and antioxidants interactions based on Lanchester's equation and its validation. Section 4 details the computational model in C++. At last section 5 presents the work conclusions.

## II. FREE RADICALS AND ANTIOXIDANTS

Reactive oxygen species (ROS) are being continuously produced in living organisms. These molecules play several physiological functions, like signaling pathways activation in normal cells in order to modulate inflammation, regulation of smooth muscle tone and apoptosis (Ling, 1996)

These molecules are a special kind of free radicals, however in wide sense this expression defines oxygen based atoms, molecules or ions characterized by the presence of unpaired electrons (Halliwell, 1989). This feature explains their high instability and extremely chemically reactive and very short-lived.

The oxygen composition of free radicals becomes highly toxic to living cells. This event occurs in taking electrons from other molecules. The most toxic of them is the Hydroxyl radicals (.OH), which is far less abundant and highly reactive and considered as the responsible for most of the oxidative damage in aerobic cells (Jomova et al, 2010), (Ren et al, 2010).

On the other side, antioxidants are molecules capable of inhibiting or delaying chemical reactions of oxidation of other molecules. It basically occurs by giving electrons to stabilize free radicals and thereby stopping chain reactions. This capability is approached by cells to counteract these damaging effects of ROS.

## III. BIOLOGICAL MODELS

### A. Mathematical modeling

The mathematical model is based on Lanchester's classical differential equation which describes the number of survivors in a conventional confrontation (Jiuyong et al, 2011). This equation represents two forces coming across in a combat; both factions are composed of homogeneous forces. The model provides a prediction of combat time and how it will be structured by changes in given parameters.

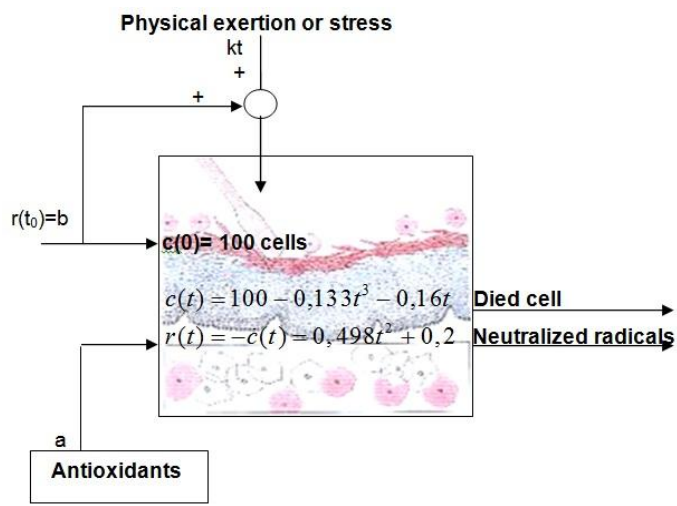

Fig. 1: Simplified model of free radicals and antioxidants dynamics

---

[1] Alvaro Juan Ojeda is with the IES Ferran Tallada, Gran Vista 54, 08032 Barcelona, Spain email: alvaro.juan4321@gmail.com

The proposed model describes the number of surviving cells in cell culture c (t), after being affected by the action of free radicals r (t) over a determined time horizon. This involvement is partly offset by the protective action of antioxidants a(t). The mathematical expression represents free radicals attacking a cell culture (see equation 1 and 2), which are previously coated by antioxidants.

$$\frac{dc(t)}{dt} = -\alpha \cdot r(t) \quad (1)$$

$$\frac{dr(t)}{dt} = kt \quad (2)$$

Where α is the effectiveness of radicals attacks. This factor is given by "a" and "b" which respectively represent the antioxidants and initial free radicals composition within the cell culture. The effectiveness can be 100% in absence of antioxidants and then decreases so greater is the amount of antioxidants, and "k" is the production ratio of free radicals.

$$\alpha = \frac{b}{a+b} \quad (3)$$

The free radicals production is given by f(t)=kt. This function is based on Jignxian Li's related work, who studied free radicals increases under physical exertion and stress (Li et al, 1998). In order to determine the function of remaining radicals within the cell culture, the expression (1) is integrated using initial conditions.

$$r(t) = \int kt \, dt = k\frac{t^2}{2} + b \quad (4)$$

$$r(0) = b = c \quad (5)$$

$$r(t) = k\frac{t^2}{2} + b \quad (6)$$

Then deriving the equation (1) to find a relation to (2) and assuming α=0, 8 (according to Anne Hanneken's work):

$$\frac{d^2c(t)}{dt^2} = -\alpha\frac{dr(t)}{dt} = -0,8\frac{dr(t)}{dt} \quad (7)$$

The constant "k" is preliminarily assumed equal to 1. This value indicates a positive slope of the radical production function according to the experimental work of Li. The value of "b" (initial percentage of free radicals in cell culture) is assumed as 0.2 (20%) is:

$$\frac{d^2c(t)}{dt^2} = -0,8(1t) = -0,8t \quad (8)$$

Then, solving the differential equation, surviving cell function becomes:

$$c(t) = -0,133t^3 + c_1 t + c_2 \quad (9)$$

Further to find the values of the constants c1 and c2 are used initial conditions c'(0) = -αb and c(0) = 100. The values represent the amount of accumulation of radicals and initial cell number of cells at the initial time. Solving this equation results:

$$c(t) = 100 - 0,133t^3 - 0,8(0,2)t \quad (10)$$

$$c(t) = 100 - 0,133t^3 - 0,16t \quad (11)$$

This function is depicted in figure 2:

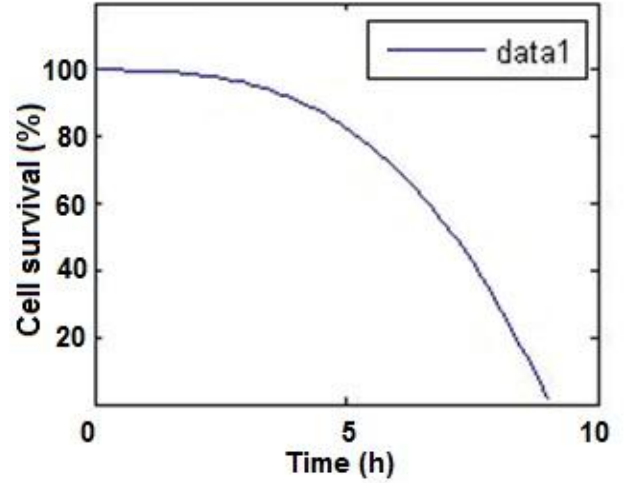

Fig. 2: Cell survival function

To find the free radical function r (t) the equation (1) is derived:

$$r(t) = -\frac{1}{\alpha}(\frac{dc(t)}{dt}) \quad (12)$$

$$r(t) = -(\frac{1}{0,8})(-0,4t^2 - 0,16) \quad (13)$$

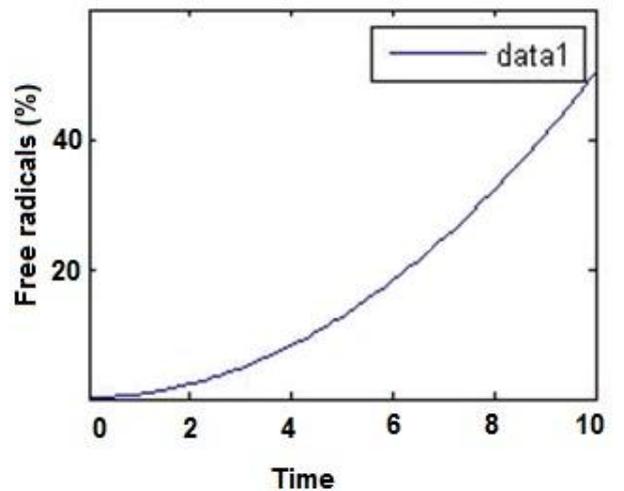

Fig. 3: Free radical's accumulation function

B. *Model validation*

The mathematical model was compared to Anne Hanneken's work in dealing with human retinal pigment epithelial - RPE cells. In such work cell cultures were treated with either hydrogen peroxide (H2O2) or t-butyl hydroperoxide (t-BOOH) (free radicals).

The MIT assay methods were considered to evaluate ability of specific flavonoids (antioxidants) to protect RPE cells from cell death. The protection efficacy was in range from 80% to 100%. These results were taken of reference to determine a typical behavior of cells exposed to oxidative-stress–induced death. The proposed mathematical function describes closely this behavior in different scenarios.

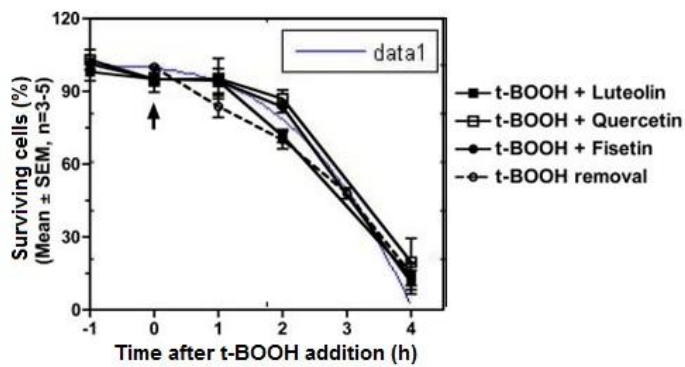

Fig. 4: Model validation

To understand the behavior of this dynamic, the function of surviving cells is represented in different scenarios, by changing whether efficiency (α) constant of free radicals production (k). By lower attack efficiency of free radicals the survivor cells prolong their resistance, otherwise accelerate their dead time (see figure 5). On the other hand, by lower free radicals production (given by the function slope) the survivor cells live longer time (see figure 6)

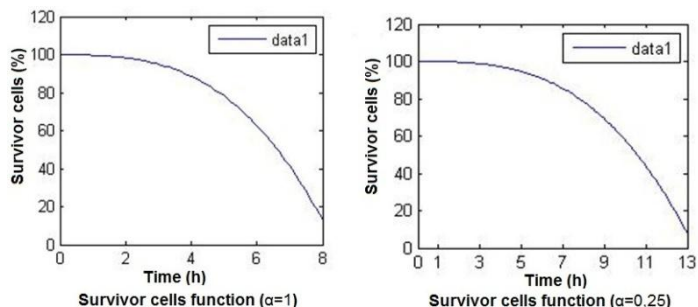

Fig. 5: Sensitivity analysis modifying the amount of antioxidants

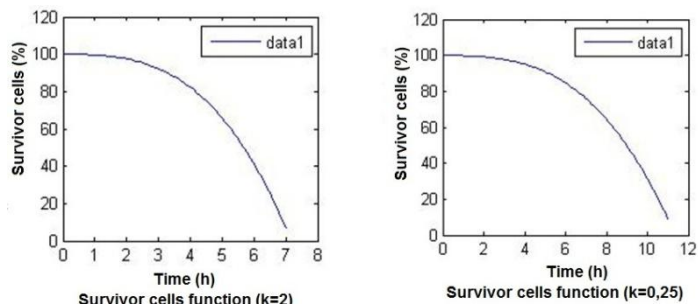

Fig. 6: Sensitivity analysis modifying free radicals production

## IV. COMPUTATIONAL MODEL

This model represents skin apoptosis due to free radicals excess, that is to say, death cell produced by physical exertions or stressful situations. The model presents two inputs: excess of free radicals production (r) and antioxidants (a). The program calculates dead cell when the number of free radicals are greater than antioxidants.

The algorithm is based on the network theory of aging. This theory proposes that the aging is caused by defective mitochondria, aberrant proteins and free radicals production. (Kirkwood and Kowald, 1997).

This program calculates dead cells of an individual, who takes part in social activities, linked to energy consumption under a daily timetable. In order to represent this biological process, the basic expression for summation of free radicals is given by equation (14) and (15). The program report is graphically depicted in figure 7.

$$\frac{n(n+1)}{2} = 30000 \longrightarrow n(n+1) = 60000 \quad (14)$$

$$n^2 + n - 60000 = 0 \quad (15)$$

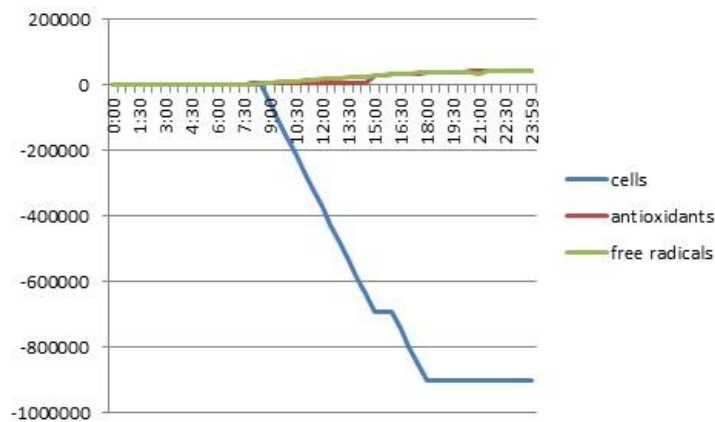

Fig.7: Graphic of simulation program

## V. CONCLUSIONS

In order to adequately explain dynamics of complex biological systems, mathematical and computational models were developed to describe and represent free radicals and antioxidants dynamics in cells environment. The mathematical model depicts this interaction in a cell culture and the C++ based simulation program in a human organism.

The Lanchester's differential equation is suitable to describe inner mechanism between free radicals and antioxidants within a cell culture. The function of free radicals production after physical effort and stress was helpful to achieve this mathematical model.

This function was contrasted with Anne Hanneken's experimental work from the Department of Molecular and Experimental Medicine, published by the journal *Investigative Ophthalmology and Visual Science* in July, 2006. The mathematical model closely describes the behavior of survivor cells obtained in this experiment.

On the other hand, the simulation program correctly calculates the amount of dead cells in the skin of an individual due to an excess of free radicals, that is to say, around 30000 to 40000 skin cells, quantity that agrees with cited bibliography. When antioxidants have diminished, free radicals attack cells producing apoptosis.

Both model models represent the same biological process but in different scenarios: the mathematical model represents this process in a cell culture and the computational in an individual's organism. The two models complement each other: the mathematical models are adequate for systems with few variables; while computational model can approximate better these mechanisms when this process becomes more complex.

**ACKNOWLEDGMENT**:

Thanks to God for his blessings along this work, thanks also to Dr. Sílvia Maymó and Dr. Jordi Cortadella for their corrections.

**BIBLIOGRAPHY**:

- **Edwin Wang, A roadmap of cancer systems biology. Publisher: Nature Publishing Group, issue: 713, pp. 1-28, 2010.**
- **Q. Cui, E. Purisima and E. Wang, Protein evolution on a human signaling network BMC Syst. Biol, vol 3, n 21, 2009.**
- **Taylor, R. Linding, D. Warde-Farley et al, Dynamic modularity in protein interaction networks predicts breast cancer outcome Eighth International Conference on Fuzzy Systems and Knowledge Discovery(FSKD), 2011.**
- **Z. Jiuyong, X. Chuanqing, G. Lei and D. Yuan, The mathematical model based on the battle of Berlin Nature Biotechnol 27, pp. 199-204, 2009.**
- **Hanneken, F. Lin, J. Johnson and P. Maher, Flavonoids protect human retinal pigment epithelial cells from oxidative-stress-induced death. Invest. Ophthalmol. Vis. Sci. vol. 47, n 7, pp. 3164-3177, 2006.**
- **Low Ling, In full bloom: look fabulous during and after pregnancy. Marshall Cavendish Editions, 2007.**
- **Low Ling, Reactive oxygen species and programmed cell death. Trends Biochem Sci, vol. 21(3), pp. 83-89, 1996.**
- **B. Halliwell and J. Gutteridge, Free Radicals in Biology (2th ed.). Oxford Press University, 1989.**
- **K. Jomova, D. Vondrakova, M. Lawson and M. Valko, Metals, oxidative stress and neurodegenerative disorders. Molecular and Cellular Biochemistry, vol. 345 (1 - 2), pp. 91-104, 2010.**
- **J. Ren, L. Pulakat, A. Whalley-Conell and J. Sowers, Mitochondrial biogenesis in the metabolic syndrome and cardiovascular disease. Journal of Molecular Medicine, vol. 88 (10), pp. 993 - 1001, 2010.**
- **Zhang Jiuyong, Xu Chuanqing, Gong Lei and Dehui Yuan, The mathematical model based on the battle of Berlin. Eighth International Conference on Fuzzy Systems and Knowledge Discovery (FSKD), pp. 2133-2136, 2011.**
- **J. Li, J. Chan, K. LU, J. Xin, D. Li and H. Gao, Effect of verbascoside on reducing the concentration of oxygen free radicals of skeletal muscle during and after exercise. Proceedings of 20th annual International Conference of the IEEE Engineering in Medicine and Biology Society, 1998.**
- **T. Kirkwood and A. Kowald, Network theory of aging. Experimental Gerontology, vol. 32, pp. 395 - 399, 1997.**